\documentclass{article}
\usepackage{spconf,amsmath,graphicx, kotex, hyperref}
\usepackage{array}

\def\paireddata{\texttt{\{Pop, Piano Cover\}}}
\def \ourmodel{Pop2Piano$_{\text{PSP}}$}
\def \baselinemodel{Pop2Piano$_{\text{POP909}}$} 

\newcommand{\figref}[1]{Fig \ref{#1}}

\newcolumntype{?}{!{\vrule width 1pt}}
\title{Pop2Piano : Pop Audio-based Piano Cover Generation}
\twoauthors
 {Jongho Choi\sthanks{The first author performed the work
	while at Music and Audio Research Group, Seoul National University.}}
	{Rebellions Inc.\\
	\tt{jongho.choi@rebellions.ai}\\
 \\
        }
 {Kyogu Lee}	{Department of Intelligence and Information,\\
        Interdisciplinary Program in Artificial Intelligence,\\
        AI Institute, Seoul National University\\
	\tt{kglee@snu.ac.kr}}
\begin{document}
\ninept
\maketitle
\begin{abstract}
Piano covers of pop music are enjoyed by many people. However, the task of automatically generating piano covers of pop music is still understudied. This is partly due to the lack of synchronized \paireddata{} data pairs, which made it challenging to apply the latest data-intensive deep learning-based methods. To leverage the power of the data-driven approach, we make a large amount of paired and synchronized \paireddata{} data using an automated pipeline. In this paper, we present Pop2Piano, a Transformer network that generates piano covers given waveforms of pop music. To the best of our knowledge, this is the first model to generate a piano cover directly from pop audio without using melody and chord extraction modules. We show that Pop2Piano, trained with our dataset, is capable of producing plausible piano covers. 

\end{abstract}

\let\thefootnote\relax\footnotetext{This work was supported by Institute of Information \& communications Technology Planning \& Evaluation (IITP) grant funded by the Korea government(MSIT) (No. 2022-0-00320, Artificial intelligence research about cross-modal dialogue modeling for one-on-one multi-modal interactions)}

\section{Introduction}
\label{sec:intro}

\textit{Piano cover} refers to a musical piece in which all musical elements of an existing song are recreated or arranged in the form of a piano performance. Piano covers of pop music are one of the widely enjoyed forms of music. For example, people use them for music education purposes, and piano cover creators have millions of subscribers on media such as Youtube.

In order for a human to create a piano cover, it is necessary to recognize all musical elements such as melodies, chords or moods from the original audio and reinterpret them into musically appropriate piano performances. Therefore, making a piano cover is not an easy task even for humans as it is a creative task and requires musical knowledge.

Previous studies have focused on arranging given pop audio to other instruments \cite{ariga2017song2guitar, takamori2019audio} using multiple external modules to extract explicit musical information, such as melodies and chords. However, creating a piano cover involves various implicit musical characteristics, such as the atmosphere of the music and the arranger's style. Therefore, we believe that research on end-to-end conversion between audio and piano covers can also be valuable.

Meanwhile, Deep learning has demonstrated excellent performance in modeling high-dimensional music audio data. However, to the best of our knowledge, there has been no deep learning study on piano cover modeling using waveforms directly from pop music. This may be due to the lack of large amounts of synchronized paired data for such modeling.

In this study, we introduce a Transformer-based piano cover generation model, called Pop2Piano, which generates a piano performance (MIDI) from the waveform of pop music. Our contributions are summarized as follows:

\begin{itemize}
    \item We propose a novel approach to piano cover generation. For this, We build 300-hour of synchronized \paireddata{} dataset, called \textbf{P}iano Cover \textbf{S}ynchronized to \textbf{P}op Audio (PSP), and introduce the preprocessing system used to make this dataset.
    \item We design Pop2Piano, a Transformer network that generates piano covers from given audio. It does not use a melody or chord extractor but uses the spectrogram itself as an input. It also controls the covering style using the arranger token.
    \item We upload a list of data and preprocessing codes to reproduce the PSP dataset, and release an executable Pop2Piano demo on Colab\footnote{\href{https://sweetcocoa.github.io/pop2piano_samples}{https://sweetcocoa.github.io/pop2piano\_samples}}.
\end{itemize}

\begin{figure*}[htb]
  \centering
  \centerline{\includegraphics[width=0.9\textwidth]{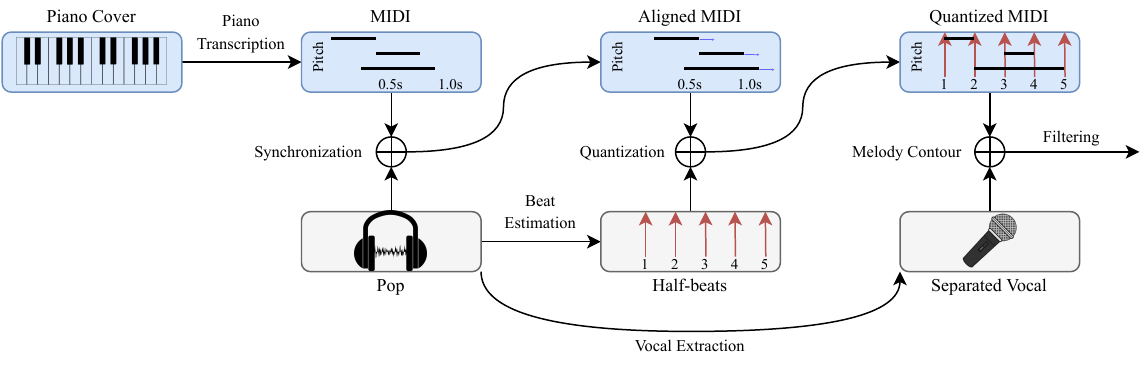}}
\caption{A preprocessing pipeline for synchronizing and filtering paired \paireddata{} audio data.}
\label{fig:datapipeline}
\end{figure*}

\section{Related Work}
\label{sec:relatedwork}
\subsection{Automatic Music Transcription}
\label{subsec:amt}

Automatic music transcription (AMT) is the task of estimating note information from musical instrument waveforms. In piano AMT, several studies have used CNN and RNN architectures to extract features and model note onset, offset, and frame \cite{hawthorne2017onsets, kim2019adversarial, kong2021high}. Some studies have employed the Transformer model and spectrograms to generate MIDI \cite{hawthorne2021sequence, gardner2021mt3}. Multi-instrument AMT aims to estimate note information for each instrument from mixed sounds of several instruments. Cerberus \cite{manilow2020simultaneous} and MT3 \cite{gardner2021mt3} are two such studies, with RNN and Transformer models, respectively.

These studies have yielded promising results with large synchronized instrument-audio datasets, such as MAESTRO \cite{hawthorne2018enabling} and Slakh2100 \cite{manilow2019cutting}. To train a model for piano cover generation, we build a 300-hour synchronized piano cover dataset. While AMT and piano cover generation share similarities in converting audio to musical notes, they differ in that piano cover generation does not have a single correct answer. For example, the monophonic vocal of the original song can be expressed in various ways, such as voicing or doubling, and the harmony may appear as various accompaniment textures depending on the original song's atmosphere or arranger's style.

\subsection{Music Transformation}

Music transformation is an active research area aimed at manipulating music data. One common approach is music style transfer, which is the task of transforming a music piece from one style to another while preserving the original musical characteristics \cite{cifka2020groove2groove, https://doi.org/10.48550/arxiv.1912.05537, lu2018transferring}. 

In many studies, there is no unified definition of music style, so in this study, we use the term style to describe the musical interpretation and composition process that varies from arranger to arranger when creating a piano cover of a given pop song. There are several studies on music style transfer in this sense \cite{https://doi.org/10.48550/arxiv.1912.05537, musenet2019} 

Another common approach is music reduction, which is the process of rearranging a piece of music to be performed by a smaller group of instruments or a single performer, while preserving its musical content \cite{chiu2009automatic, onuma2010piano}. Studies on audio-based instrument covers can be considered as music reduction to a single instrument. Previous works have utilized external modules to extract melodies and chords from audio. For example, Song2Guitar \cite{ariga2017song2guitar} generated guitar covers from pop audio by using modules to extract melodies, chords, and beats. Then, a guitar tab score was generated by statistically modeling the fingering probability. Similarly, in \textit{Takamori et al.} \cite{takamori2019audio}, melodies, chords, and choruses were extracted using external modules. A piano score was then generated using rule-based methods. In contrast, our study, Pop2Piano, only uses a beat extractor to generate piano notes directly from pop audio.

\begin{figure*}[htb]
  \centering
  \centerline{\includegraphics[width=0.95\textwidth]{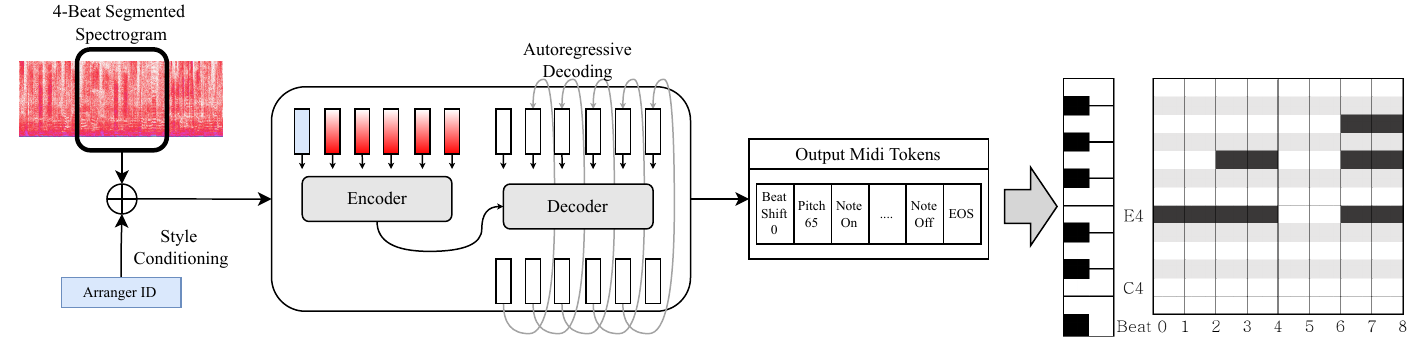}}
\caption{The architecture of our model is an encoder-decoder Transformer. Each input position for the encoder is one frame of the spectrogram. We concatenated an embedding vector representing a target arranger style to the spectrogram. Output MIDI tokens are autoregressively generated from the decoder.}
\label{fig:training}
\end{figure*}

\section{Dataset}
\label{sec:dataset}
The main bottleneck in modeling raw waveform pairs is a long range of dependencies, as they are computationally challenging to learn the high-dimensional semantics of music. An alternative to that is a method of training short-length waveform segments in pairs. However, this method requires paired data to be synchronized; otherwise, there is a risk that the correct label will not be included in the paired segmented data. As the collected data pair was not synchronized, we developed a preprocessing pipeline to obtain synchronized \paireddata{} data, as shown in \figref{fig:datapipeline}.

\subsection{Preprocessing}

\subsubsection{Synchronizing Paired Music}
First, we convert piano cover audio into MIDI, using the piano transcription model \cite{kong2021high}. Second, we roughly align piano MIDI with pop audio. We obtain a warping path using \emph{SynctoolBox} \cite{Mueller2021} and then use it to modify the note timings of MIDI via linear interpolation. This provides a simple synchronization between the piano cover and pop audio. Third, the note timings are quantized into 8th-note units. After extracting beats from pop audio using \emph{Essentia}\footnote{https://essentia.upf.edu/}, the onset and offset of each note in MIDI are quantized to the nearest 8th-note beat. If the onsets and offsets of the quantized notes are the same, the offset is moved to the next beat. In this way, the entropy of data can be lowered by changing the time unit of a note from continuous-time (seconds) to quantized time (beats).

\subsubsection{Filtering Low Quality Samples} \label{subsubsec:filtering}

There are cases where the data pairs are unsuitable for various reasons, such as the difference in musical progress or having different keys from the original song. To filter out these cases, all data with a melody chroma accuracy (MCA) \cite{raffel2014mir_eval} of 0.15 or less, or an audio length difference of 20\% or more are discarded. Melody chroma accuracy is calculated between the pitch contour of the vocal signal extracted from the audio and the top line of the MIDI. 

We use \emph{Spleeter} \cite{spleeter2020} to separate the vocal signal. Then, to get the melody contours of pop music, the f0 sequence of the vocal is calculated using \emph{Librosa} \cite{brian_mcfee_2022_6097378} \emph{pYIN} \cite{6853678}. The sample rate is 44100 and the hop length is 1024.

\subsection{Piano Cover Synchronized to Pop Audio (PSP)}
We collect 5989 piano covers from 21 arrangers and corresponding pop songs on YouTube. We then synchronized and filter \paireddata{}. As a result, a total of 4989 tracks (307 hours) are left, which are used as a training set, PSP. Note that the piano cover included in the PSP is unique, but the original song is not. This will help the model train separately the style of the piano cover according to the arranger conditions and the acoustic characteristics of the given audio.

\section{Model}

\subsection{Inputs and Outputs}
Pop2Piano uses a log-Mel spectrogram of pop audio as an encoder input. The sampling rate is 22050, the window size is 4096, and the hop size is 1024. In addition, the arranger token, which indicates who arranged the target piano cover, is embedded and appended before the first frame of the spectrogram. Each step of the decoder output is chosen from the following types of tokens: 

\begin{description}
    \item[Note Pitch] [128 values] Indicates a pitch event for one of the MIDI pitches. But only the 88 pitches corresponding to piano keys are actually used.
    \item[Note On/Off] [2 values] Determines whether previous Note Pitch events are interpreted as note-on or note-off.
    \item[Beat Shift] [100 values] Indicates the relative time shift within the segment quantized into 8th-note beats. It will apply to all subsequent note-related events until the next Beat Shift event. We define the vocabulary with Beat Shifts up to 50 beats, but because time resets for each segment, In practice we use only about 10 events of this type.
    \item[EOS, PAD] [2 values] Indicates the end of the sequence and the padding of the sequence.
\end{description}

\begin{figure}[ht]
  \centerline{\includegraphics[width=\linewidth]{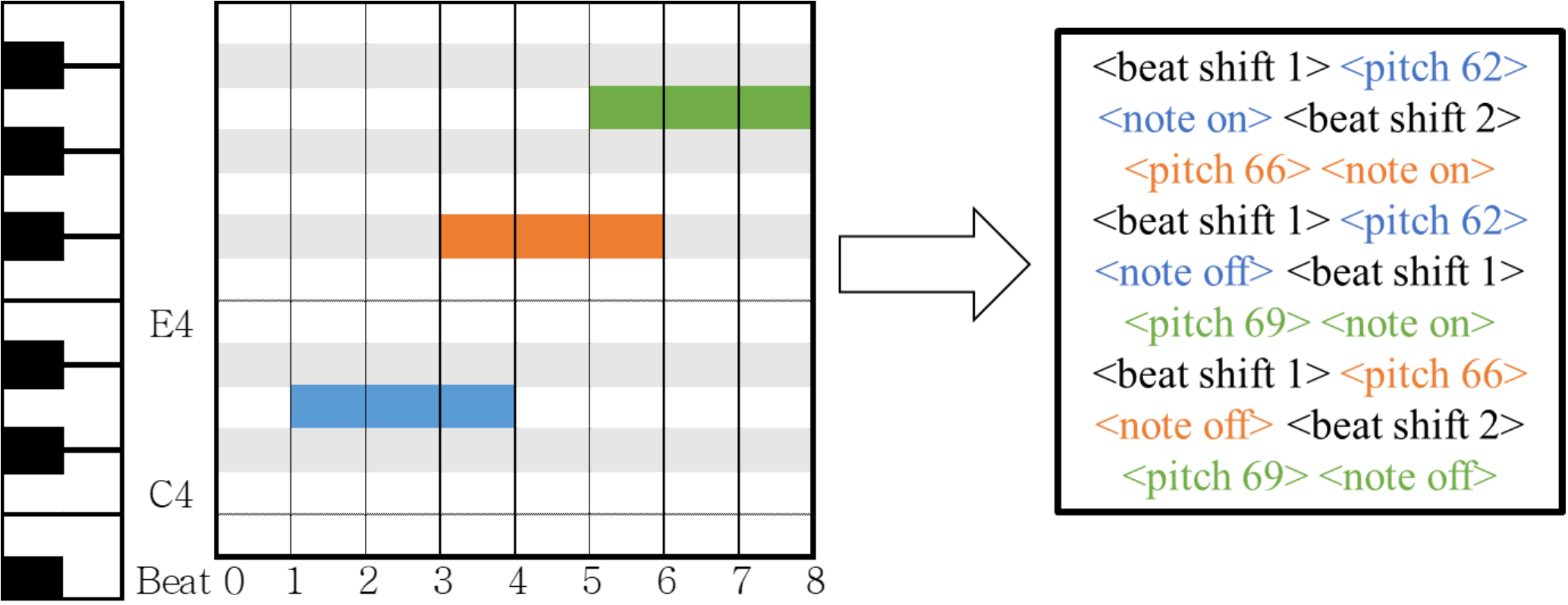}}
  \caption{An example of piano tokenization. the beat shift token means a relative time shift from that point in time.}
\label{fig:midi_like}
\end{figure}

The idea for this dictionary was taken from Transformer-based AMT studies \cite{hawthorne2021sequence, gardner2021mt3}. For a detailed example of this token's interpretation, See \figref{fig:midi_like}. For each input, the model autoregressively generates outputs until an EOS token is generated in the decoder. To generate a piano cover of pop audio of arbitrary length, in the inference stage, the audio is sequentially cropped by 4 beats and used as input to the model, then generated tokens (except for EOS) are concatenated. After that, the relative beats of the generated tokens are converted into absolute time using the absolute time information of the beat extracted from the original song and then converted into a standard MIDI file. 

\subsection{Architecture}
The Pop2Piano model architecture is T5-small \cite{raffel2019exploring} used for MT3 \cite{gardner2021mt3}. It is a Transformer network with an encoder-decoder structure. The number of learnable parameters is about 59M. Unlike MT3 \cite{gardner2021mt3}, the relative positional embedding of the original T5 is used instead of the absolute positional embedding. Additionally, a learnable embedding layer is used for embedding the arranger style. An overview of these processes is shown in \figref{fig:training}.

\section{Experiment}
\label{sec:experiment}

In this experiment, we check whether Pop2Piano trained with PSP (\ourmodel{}) can generate a plausible piano cover and verify that it follows a specific arranger's style.

\subsection{Training Setup}
We train the model with audio extracted at random 4-beat lengths. We repeat this process 2000 times for the entire PSP dataset. Assuming that the average music has a bpm of 120, we can estimate that about 5500 hours of audio were used for training using the PSP dataset. The network is optimized using the AdaFactor \cite{shazeer2018adafactor} and its learning rate is 0.001. 

\subsection{Generation Result}
\begin{figure}[ht]
\begin{minipage}[b]{.48\linewidth}
  \centering
  \centerline{\includegraphics[width=\linewidth]{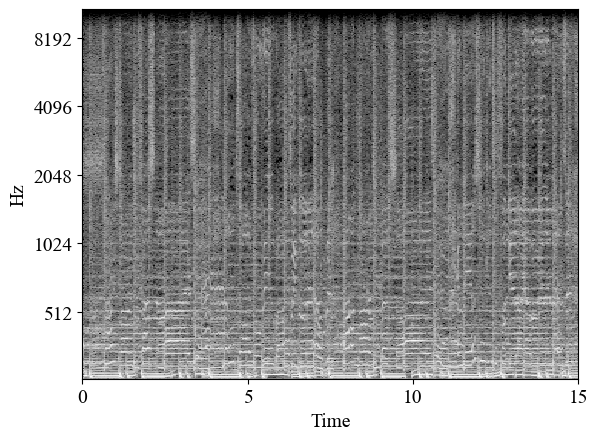}}
  \centerline{(a)}
\end{minipage}
\begin{minipage}[b]{0.48\linewidth}
  \centering
  \centerline{\includegraphics[width=\linewidth]{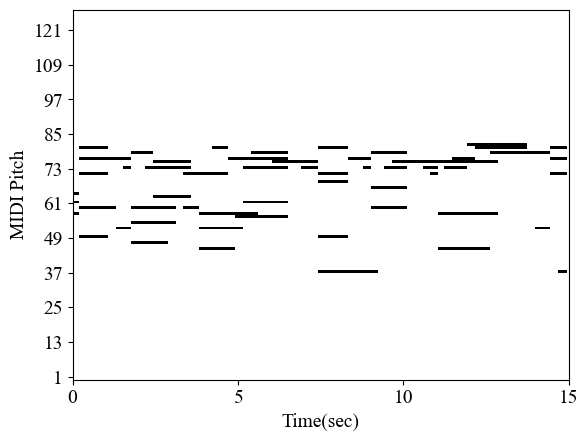}}
  \centerline{(b)}
\end{minipage}
\caption{(a) A Melspectrogram of an input pop audio. (b) A piano roll of output MIDI notes corresponding to input audio.}
\label{fig:gen}
\end{figure}

\begin{figure}[htb]
  \centering
  \centerline{\includegraphics[width=.8\linewidth]{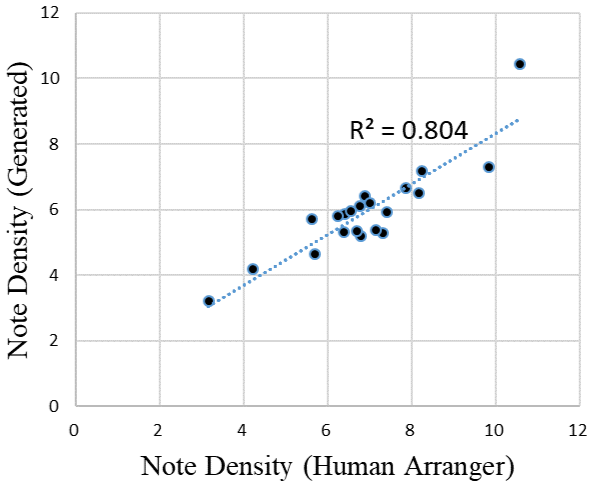}}
\caption{The dots represent the note density of the piano cover generated by being conditioned with that arranger. The unit is the number of notes per second.}
\label{fig:notedensity}
\end{figure}

\figref{fig:gen} is an example of a piano cover generated using arbitrary pop music as an input. The generated sample can be listened to on the link\footnote{\href{https://sweetcocoa.github.io/pop2piano_samples/fig_on_paper.html}{https://sweetcocoa.github.io/pop2piano\_samples/fig\_on\_paper.html}}. The original song is a complex audio signal mixed with various sounds such as vocals, bass, and percussion. Nevertheless, it can be seen that the generated piano cover shows the vocal melody line of the original song as stacked notes, and also includes a plausible accompaniment that follows the harmony of the original song. Also, various piano cover samples generated by Pop2piano can be listened to with the original song on the demo page.



Additionally, we would like to confirm whether the generated piano cover not only plausibly followed the original song but also covered it in the style of the arranger we designated. However, the cover style is very implicit, so there are few ways to evaluate quantitatively. Inspired by the difference in note densities among arrangers, we measure the note densities of the piano cover generated by \ourmodel{} to indirectly verify whether it generates a piano cover in the target arranger's style. As a result, \figref{fig:notedensity} shows that the generated piano cover follows the note density of the target arranger with high linearity of $R^2=0.804$.


\begin{table}
 \begin{center}
 \begin{tabular}{|l|l|l|}
  \hline
  Original Songs & Arranger & Average MCA \\
  \hline
  POP909$_F$ & Human  & 0.395 \\
  \hline
  PSP & Human & 0.3493 \\
  \hline
  \hline
  POP909$_F$ & \ourmodel{}  & \textbf{0.402 $\pm$ 0.021} \\
  \hline
 \end{tabular}
\end{center}
 \caption{Average Melody Chroma Accuracy (AMCA) of human-made and generated piano covers. The piano covers of Pop2Piano are generated for all 21 arrangers of PSP, and its standard deviation according to arrangers are shown in the table.}
 \label{tab:chroma_acuracy}
\end{table}

We assess the quality of the generated piano covers by computing the MCA, as described in Section \ref{subsubsec:filtering}. Interestingly, our results show that the MCA of the generated piano covers is similar to that of the human-created POP909 dataset \cite{wang2020pop909}, even when trained without explicit melodies. POP909 is a piano MIDI dataset of 909 Chinese pop songs, with separate tracks for melody, bridge, and accompaniment. However, we merge all the tracks for consistency in our experiments, and like PSP, we synchronize the \paireddata{} pairs through our preprocessing pipeline. After filtering, 817 tracks remain, which we denote as POP909$_F$ in the rest of this paper. The MCA values are presented in Table \ref{tab:chroma_acuracy}. Note that a higher MCA does not necessarily indicate better piano covers, and thus, evaluating the quality of the generated piano covers will also be a good follow-up study.

\subsection{Subjective Evaluation}
We measure the subjective quality of the generated piano covers by evaluating their naturalness. Since there is no prior study publicly available, we use a Pop2Piano network trained with POP909 (\baselinemodel{}) as a baseline model. The comparison with this model indicates how the PSP dataset affected the performance of the Pop2Piano model.

We conduct a user study. There are 25 participants and no professional musicians are included. Participants do not know which model has made each piano cover. We select 10-seconds from the beginning of the chorus of 25 songs that are not included in the training dataset. For each song, a piano cover made by a human arranger is used as GT. However, since GT also needs to be synchronized with the original song, the synchronization pipeline is applied and that data is denoted as GT$_F$. After listening to the original song and piano covers, the participants evaluate how naturally the given piano cover arranged the original song with 1-5 points.

In the listening evaluation, 70\% of participants prefer the piano cover generated by \ourmodel{} to \baselinemodel{}. In MOS analysis between \ourmodel{} and \baselinemodel{}, using paired one-sided Wilcoxon test, we reject H$_0$ at 99\% confidence intervals $(0.222, \texttt{inf})$ with $p=3.34e$$-05$. See Table \ref{tab:preference}.

\begin{table}
 \begin{center}
 \begin{tabular}{|l|c?c|c|}
  \hline
  (\%) & GT$_F$ & \ourmodel{} & \baselinemodel{} \\
  \hline
  \hline
  vs GT$_F$ & - & \textbf{29.6} & 22.24\\
     \hline
  vs \ourmodel{} & \textit{70.4} & - & 36 \\
  \hline
  vs \baselinemodel{} & \textit{77.76} & \textbf{64} & - \\
  \hline
  \hline
  MOS & \textit{3.771} & \textbf{3.216} & 2.856\\
  \hline
 \end{tabular}
\end{center}
 \caption{The winning rate and Mean Opinion Score(MOS) according to the piano cover model. GT$_F$ denotes synchronized piano covers created by human arrangers.}
 \label{tab:preference}
\end{table}

\subsection{Limitation}
We recognize that some improvements can be made to our model. For instance, Pop2Piano uses only four-beat length audio for the context of input. Therefore, features such as melody contour or texture of accompaniment have less consistency when generating longer than four-beat. Also, time quantization based on eighth note beats prevents the model from generating piano covers with other rhythms such as triplets, 16th notes, and trills.

\section{Conclusion}
\label{sec:conclusion}
We present Pop2Piano, a novel study on generating pop piano covers directly from audio without using melody or chord extraction modules based on a Transformer network. We collect PSP, 300 hours of paired \paireddata{} datasets, to train the model. And we design a pipeline that synchronizes them in a form suitable for training neural networks. We open-source the list of data and code needed to reproduce the PSP. We show and evaluate that \ourmodel{} can generate plausible pop piano covers and also can mimic the style of a specific arranger.



\vfill\pagebreak

\bibliographystyle{IEEEbib}
\bibliography{strings,refs}

\end{document}